\documentclass[conference]{IEEEtran}
\IEEEoverridecommandlockouts
\usepackage[whole]{bxcjkjatype}
\usepackage{cite}
\usepackage{amsmath,amssymb,amsfonts}
\usepackage{algorithmic}
\usepackage{graphicx}
\usepackage{textcomp}
\usepackage{xcolor}

\usepackage{slashbox}
\usepackage{comment}
\usepackage[caption=false]{subfig}

\usepackage[T1]{fontenc}
\usepackage{lmodern}

\def\BibTeX{{\rm B\kern-.05em{\sc i\kern-.025em b}\kern-.08em
    T\kern-.1667em\lower.7ex\hbox{E}\kern-.125emX}}
\begin{document}

\title{Analysis of Short Dwell Time in Relation to User Interest in a News Application}

\author{
    \IEEEauthorblockN{Ryosuke Homma\IEEEauthorrefmark{1} \quad Yoshifumi Seki\IEEEauthorrefmark{2} \quad Mitsuo Yoshida\IEEEauthorrefmark{1} \quad Kyoji Umemura\IEEEauthorrefmark{1}}
    \IEEEauthorblockA{\IEEEauthorrefmark{1}\textit{Toyohashi University of Technology}\\
    Aichi, Japan\\
    homma.ryosuke.vt@tut.jp, yoshida@cs.tut.ac.jp, umemura@tut.jp}
    \IEEEauthorblockA{\IEEEauthorrefmark{2}\textit{Gunosy Inc.}\\
    Tokyo, Japan\\
    yoshifumi.seki@gunosy.com}
}

\maketitle

\begin{abstract}
Dwell time has been widely used in various fields to evaluate content quality and user engagement.
Although many studies shown that content with long dwell time is good quality, contents with short dwell time have not been discussed in detail.
We hypothesize that content with short dwell time is not always low quality and does not always have low user engagement, but is instead related to user interest.
The purpose of this study is to clarify the meanings of short dwell time browsing in mobile news application.
First, we analyze the relation of short dwell time to user interest using large scale user behavior logs from a mobile news application.
This analysis was conducted on a vector space based on users click histories and then users and articles were mapped in the same space.
The users with short dwell time are concentrated on a specific position in this space; thus, the length of dwell time is related to their interest.
Moreover, we also analyze the characteristics of short dwell time browsing by excluding these browses from their click histories.
Surprisingly, excluding short dwell time click history, it was found that short dwell time click history included some aspect of user interest in 30.87\% of instances where the cluster of users changed.
These findings demonstrate that short dwell time does not always indicate a low level of user engagement, but also level of user interest.

\end{abstract}

\vspace{0.2cm}


\section{Introduction}
How can we measure the extent to which users are attracted to content?
The click-through rate (CTR) is mainly used to evaluate a user interest in the content of web pages or digital advertisement recommendations.
However, clicks do not directly reflect user content preferences~\cite{lu2018between}.
One of the reasons for this is ``clickbait,'' which is content whose main purpose is to attract attention and encourage visitors to click on a link to a particular web page and this content often causes misleading and unverified information~\cite{chen2015misleading}.
Therefore, to evaluate user interest in content, many previous studies have attempted to study post-click user engagement~\cite{guo2012beyond, kim2014modeling, hassan2013beyond}.

There are explicit and implicit measures of post-click user engagement feedback.
Explicit feedback (e.g., likes, comments, and ratings) does not occur often.
There is thus a problem of sparse data in explicit feedback~\cite{yin2013silence,yi2014beyond}.
Therefore, implicit feedback (e.g., dwell time and scroll length) is often used, especially dwell time is widely used~\cite{liu2010understanding,yi2014beyond}.
In this study, we focus on dwell time.
Because it can be recorded at a low cost and can be used for various services.

In this study, we focus on the meaning of short dwell time.
Although there have been many studies about the relationship between user satisfaction and dwell time~\cite{yin2013silence}, most of them focus on long dwell time~\cite{lu2018between}, and short dwell time has not been sufficiently discussed.
Many studies have concluded that long dwell time indicates high user engagement. 
However, is it suitable to consider all short dwell times as inherently negative?
The details of what a short dwell time can mean have not been clarified, and so this question has not been answered.

We analyze short dwell time browsing in relation to user interest in a news application.
First, we construct a vector space for visualization based on user's click histories and news articles, and then we map users and articles in the same space.
The users with short dwell time were concentrated on a specific position in the vector space; thus, short dwell time is strongly related to their interest.
We analyze the characteristics of short dwell time browsing by excluding these browses from their click histories.
Surprisingly, excluding short dwell time click history, 30.87\% of user clusters are changed; thus, it is likely that short dwell time click history includes some aspects of user interest.
These findings demonstrate that short dwell time does not always indicate a low level of user engagement, but it does indicate the level of user interest.

This study used conventional methods, such as principal component analysis (PCA), k-means clustering, and word2vec, and there are no novelties in these methodology.
However our visualization, which combines these conventional methods and visualizes users and articles in same space using users' click history, provides useful knowledge; thus, we consider that our visualizing procedures are useful and novel.
Our dataset is not public due to business confidentiality, but our findings are interesting and novel and we are sure that these results will contribute various fields, such as evaluating content quality or user engagement, social science studies, and digital journalism.

In this paper, we make the following contributions:
\begin{itemize}
    \item Applying PCA space that fits user's browsing behavior to the content to reflect usage, we show that this PCA space can visualize the content by usage.
    \item Visualizing dwell time in this space, we find that there is a relationship between user interest and dwell time.
    \item We show that a short dwell time does not always indicate a low level of user engagement because some users prefer short dwell time browsing.
\end{itemize}
\label{sec:introduction}

\section{Related Work}
There have been many studies about post-click engagement, which measures how much users prefer the content after clicking it.
Previously, content had been evaluated primarily by using  CTR~\cite{hu2008collaborative,yoneda2019algorithms}.
However, users do not always like the clicked content.
This problem is seen not only in browsing through web pages but also in watching videos~\cite{covington2016deep} and playing music~\cite{wen2019leveraging}. 

Post-click engagement is evaluated using post-click feedback.
There are two types of post-click feedback: explicit feedback (such as likes, comments, and ratings) and implicit feedback (dwell time, scroll length, and eye-tracking).
Explicit feedback has been used, but there is a problem of sparse data regarding explicit feedback~\cite{yin2013silence,yi2014beyond}.
Yin et al.~\cite{yin2013silence} analyzed user behavior in a joke sharing mobile application (JokeBox), and found that, on average, a person only voted on 2 out of 100 accessed items;\footnote{To vote means to give a user's opinions of an item.} thus, they used dwell time as user feedback instead of explicit feedback for developing recommender systems.

Implicit metrics other than dwell time have also been used.
For example, eye tracking can be used to learn which parts of an article a user is looking at.
In browsing news articles, gaze information is correlated with user engagement~\cite{arapakis2014user}.
However, obtaining eye-tracking information is expensive because it requires special equipment, and it is difficult to obtain a large amount of user data.
Therefore, the correspondence between mouse cursor movement and web page gazing has also been examined~\cite{navalpakkam2013measurement}.
However, it has been reported that those relationships do not always correspond (e.g., when the mouse cursor is not moving)~\cite{chen2001can}.
In addition, as these data include position information in one-second units, it can be expected that the data size will be large and difficult to use.
For the dwell time data, though, a large amount of data can be easily used.

There have been many studies about dwell time.
Yi et al.~\cite{yi2014beyond} claimed that the CTR does not capture post-click user engagement and used dwell time to measure user engagement.
They also claimed that dwell time varies by context.
For example, users spend on average less dwell time per article on mobile or tablet devices than on desktops.
Also, users on average spend less time per slideshow than per article.
Liu et al.~\cite{liu2011using} showed that dwell time differs depending on tasks in terms of information retrieval (IR). 
They predicted document usefulness to identify the threshold of dwell time and showed a maximum precision of 61.9\%.
Thus, dwell time is an useful metric to evaluate content quality and user engagement.

Some studies have defined a threshold of dwell time to evaluate content quality.
Lagun and Lalmas~\cite{lagun2016understanding} defined ``bounce'' as a relatively short period of disengagement by a disinterested users. 
``Bounce'' is defined as a direct return, so it is not considered a good outcome for content creators~\cite{sculley2009predicting}.
Lu et al.~\cite{lu2018between} found that users may click on a news article even if they think they dislike it, and the estimation accuracy of preferences is improved by considering a dwell time longer than a threshold (52s) to mean that the users like the article.
They have shown that dwell time above the threshold is associated with higher user engagement, but the relationships between short dwell time and user engagement have not been shown.

In a study about content with short dwell time, Seki and Yoshida~\cite{seki2018analysis} analyzed the relationship between news article categories and dwell time.
They hypothesized that a short dwell time does not necessarily indicate that the user is dissatisfied because some news articles with briefly title or image.
Further, they classify content with short dwell time by article titles and images and show a relationship between these and short dwell time.
Thus, the meaning of short dwell time has not been discussed widely.
This is the first study to attempt to discuss the relation of short dwell time and user interest in detail.
\label{sec:related_works}

\section{Data}
In this study, we studied user behavior logs and contents from Gunosy, which is a popular mobile news application in Japan.
This application provides a list of news articles, and a user taps the article to read.
In this paper, we define tapping a article cell (containing title and image) as clicking and the time between clicking and returning to the article cell list as the dwell time of the article.

We primarily analyzed user behavior logs from April 2 to April 15, 2020.
During this period, there were many news articles about the COVID-19 spreading around the world.
Users used this Gunosy for five days or more and were randomly sampled out of 30 clicks or more.
The number of users was about 12,000 and the total number of articles was about 95,000 in this period.
The total number of clicks was about 3,250,000.

These click logs include the dwell time and height of each article.
Because the height of the news article depends on the size of the device used by the user, the height of the news article is included in the click logs.
The correlations between article height and dwell time is 0.088, so this dataset has a weak correlation.
To show the robustness of our analysis, we also analyze different period logs in Section~\ref{sec:another_period}, and their details are described in that section.

\begin{figure}[tb]
  \begin{center}
  \includegraphics[width=0.75\hsize]{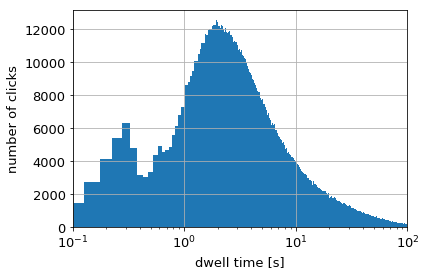}
  \caption{Distribution of dwell time in one browsing session. The dwell time peaks near 0.2s and 2s.}
  \label{fig:dwelltime}
  \end{center}
\end{figure}

Fig.~\ref{fig:dwelltime} shows a histogram of the dwell time.
There are two peaks in the dwell time distribution at around 0.2s and 2s, respectively.
It appears that it often takes around 2s for the user to determine whether to browse an article.
To analyze the characteristics of news articles with short dwell time, we defined a dwell time of 5s or less as ``short" article browsing.
This criterion is similar to that of previous studies~\cite{lagun2016understanding}.
This definition means that the first peak in Fig.~\ref{fig:dwelltime} is 2s and adds 3s as a margin.
With this threshold, 28.23\% of interactions had a short dwell time.

\label{sec:data}

\section{Visualization and Analysis}\label{sec:visualization}
\subsection{Motivation}
To analyze the characteristic of short dwell time, we visualized the article and user vectors.
In the previous study, short dwell time is was considered inherently negative, but we consider this statement to not always be true.
For example, if the purpose of clicking an article is to see an image, a user leaves immediately after seeing the image; therefore, the dwell time is short.
We hypothesize that the dwell time expresses not always the quality of the content but also the type of user interest in the content.

We visualize users and articles in the same space and analyze how articles with short dwell time are distributed in this space.
Dwell time acts as a metric that indicates the relationship between users and articles.
To understand the meaning of dwell time, we need to understand this relationship.

\subsection{Methodology}

To visualize users and articles, we adopt the vectorization method used by Gunosy and which works well as a vector for the recommender system~\cite{yoneda2019algorithms}.\footnote{Empirically, this method indicates the characteristics of Japanese news articles better than other vectorization methods.}
The article vector uses words included in the title of the article.
$W_{a}:=\left\{w_{i}\right\}_{i} \subset \mathbb{R}^{a}$, $w$ is a set of words included in the title of article $a$.
The article vector of $a$ is as follows:
\begin{equation}
a:=\frac{\sum_{w_{i} \in W_{a}} idf \left(w_{i}\right) w2v(w_{i})}{\left\|\sum_{w_{i} \in W_{a}} idf \left(w_{i}\right) w2v(w_{i})\right\|} \in \mathbb{R}^{d}.
\end{equation}
$idf(w_i)$ is the inverse document frequency obtained from past articles.
In addition, $w2v(w_i)$ is a d-dimensional vector obtained from word2vec that has trained using past news articles.

The user vector is defined by articles clicked by the user.
Let $u$ be the user and $A_u$ be the set of news articles clicked by user $u$.
Then, user vector $u$ is defined as follows:
\begin{equation}
u:=\frac{\sum_{a \in A_{u}} a}{\left\|\sum_{a \in A_{u}} a\right\|} \in \mathbb{R}^{d}.
\end{equation}
This is the average of $A_u$ article vectors. 
$d$ is set to 300 according to \cite{yoneda2019algorithms}.

To visualize article vectors and user vectors in two dimensions, we use PCA.\footnote{We considered using the non-linear method (e.g. t-SNE) for future work.}
First, we fit PCA space using user vectors.
We expect that this PCA space visualizes user interest, so we call this space {\it user interest space}. 
The article vector compressed and visualized by this PCA space represents the position of the article in the user interest space.
By comparing the position of articles with ``short dwell time'' and ``not short dwell time'' in this space, we analyze the relationship between short dwell time on user interest.

\subsection{User Interest Space}

We would like to demonstrate that our visualization method offers valid interpretations of the relationship between users and articles.
First, we analyzed the characteristics of user and article vectors on the user interest space.
We classified users and articles in clusters using k-means.
If we confirmed that the space represents user interest by visualizing users and articles, we could analyze the relationship between user interest and user dwell time.

We used k-means for 300-dimensional vectors before using PCA and set the number of clusters at four because user vectors extend in four directions in the user interest space.\footnote{Of course, we investigated various numbers of cluster; the user vector was divided neatly into four clusters.}
We used k-means++~\cite{arthur2006k} to set the initial value of the centroid.

\begin{figure}[tb]
  \begin{center}
  \includegraphics[width=0.8\hsize]{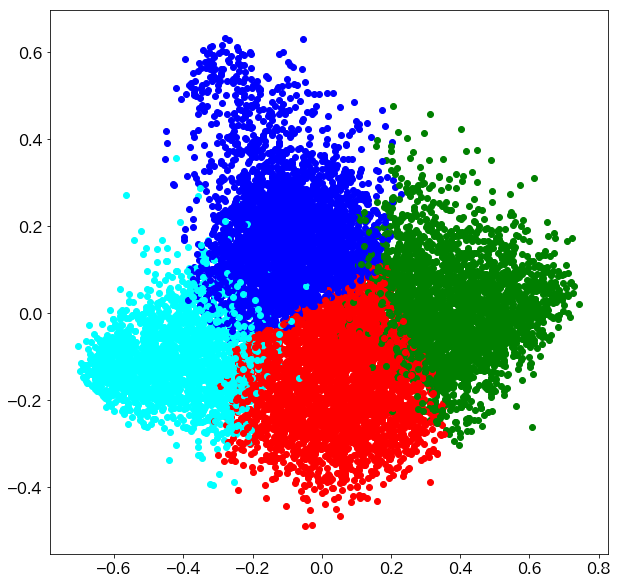}
  \caption{Visualization of the user vector (k-means, $k=4$). This space reflects user interest.
  We assigned each cluster a color based on the result of k-means before using PCA. The user vector can be neatly divided by user interest. Features appear in four directions; thus, user interest divided into four types.}
  \label{fig:user_vector}
  \end{center}
\end{figure}

Fig.~\ref{fig:user_vector} shows the user vector and its clusters.
The user vector can be neatly divided according to user interest because the clusters are located on the top, bottom, left, and right of user interest space.
This indicates that there are roughly four types of user interest.
The significance of each cluster is explained in the subsequent analysis.

\begin{figure}[tb]
    \subfloat[Fitted to user vectors]{\includegraphics[width=0.46\hsize]{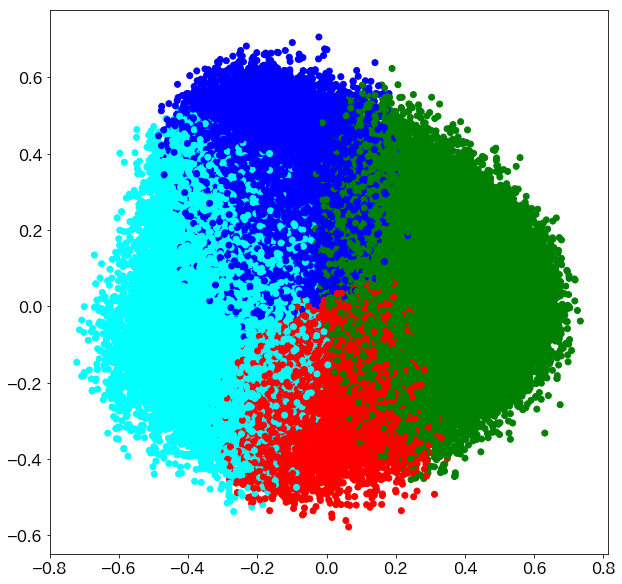}
    \label{fig:pca_user_cluster_user_plot_article}}
    \quad
    \subfloat[Fitted to article vectors]{\includegraphics[width=0.46\hsize]{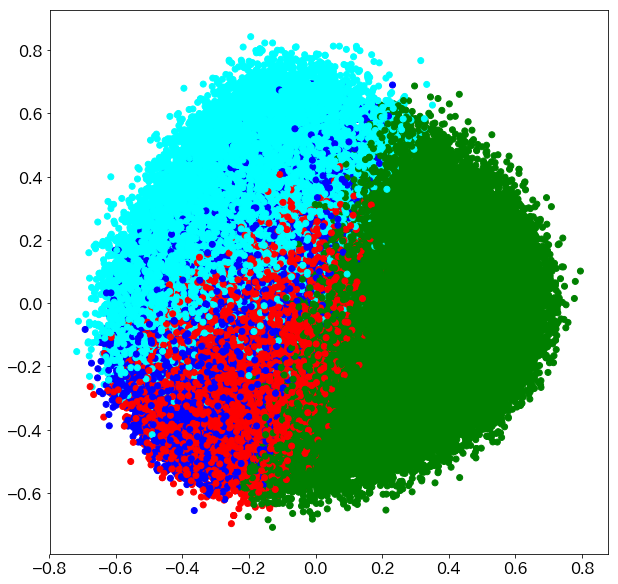}
    \label{fig:pca_article_cluster_article_plot_article}}
    \caption{Visualization of the article vector. (a) PCA: user, k-means: user, plot: article. (b) PCA: article, k-means: article, plot: article.}
    \label{fig:article_vector}
\end{figure}

Next, we compared the results of the visualization with the user interest space and the PCA space fitted to the article vector.
Fig.~\ref{fig:pca_user_cluster_user_plot_article} shows article distribution and clusters in the user interest space.
For comparison with the user interest space, Fig.~\ref{fig:pca_article_cluster_article_plot_article} shows article distribution and clusters in the PCA space fitted to the article vector. 
Two of the four clusters are not well identified because of this overlap.
By using the user interest space, we obtained a good visualization of both articles and users.
In addition, the user vector and the article vector could be compared in the user interest space.
In other words, the articles that the user clicked on and the user are located close to each other in the user interest space.

We also attempted to understand the significance of the user interest space in detail.
Table~\ref{table:cluster_words_manually} shows the regions of the user vector and the numbers of words in the article titles.
To understand the significance of user clusters, we manually defined four regions in the distribution of four user clusters.
We counted the words that appeared in the titles of articles contained in each region and interpreted the user interest that each region represented.
We defined that the position of each region corresponded to the position of user interest that each region represented.
These interpretations depended on the knowledge of the authors.
The region $R_A$ is related to ``female idol group/bikini model,'' the region $R_B$ is related to ``cooking/lifestyle,'' the region $R_C$ is related to ``entertainment,'' and the region $R_D$ is related to ``politics/domestic/COVID-19.''
In the user interest space, different news articles were included in each region, and these regions were divided according to the individuals' usage.
During the period of the data collected for this study, since the COVID-19 pandemic was prevalent around the world, $R_D$ reflects this social situation. 
If we used data in different periods, they would be divided into different regions.

\begin{table*}[tb]
  \caption{The user feature vector region and words contained in the articles.
  The following words each correspond to the instances of the word in each article. 
  The interpretation of each region depends on the knowledge of the authors.}
  \centering
  \scalebox{0.9}{
  \begin{tabular}{|r|r|r|r|r|} \hline
     Rank & $R_A$: \begin{tabular}{c}$x<=-0.4$ \\ $y<=-0.1$\end{tabular} & $R_B$: \begin{tabular}{c}$-0.4<=x<=0.0$ \\ $0.4<=y$\end{tabular} & $R_C$: \begin{tabular}{c}$ -0.2<=x<=0.2$ \\ $y<=-0.3$\end{tabular}  & $R_D$: \begin{tabular}{c}$0.45<=x$ \\ $0.0<=y$\end{tabular} \\
    & (female idol group/bikini model) & (cooking/lifestyle) & (entertainment) & (politics/domestic) \\ \hline
    1: & ファン(fan): 112 & レシピ(recipe): 330 & 志村けん(a comedian): 251 & コロナ(corona): 1582 \\
    2: & グラビア(bikini model): 98 & 人気(popularity): 256 & アナ(announcer): 156 & 新型(new type): 1132 \\
    3: & 姿(appearance): 95 & うち(house): 247 & 志村(a comedian): 149 & 緊急事態(emergency): 281 \\
    4: & 乃木坂46 (an idol group): 78 & アイテム(item): 197 & コロナ(corona): 122 & 政府(government): 253 \\
    5: & 写真集(photo book): 77 & 店(store): 183 & 大林宣彦(a director): 95 & 中国(China): 244 \\
    6: & 表紙(cover): 66 & テイクアウト(takeout): 145 & 番組(program): 78 & 感染者(infected): 188 \\
    7: & セクシー(sexy): 65 & スイーツ(sweets): 130 & 特番(special program): 62 & 新型コロナウイルス(SARS-CoV-2): 182 \\
    8: & 声(voice): 54 & ダイソー(a store): 127 & NHK (a TV station): 59 & 韓国(Korea): 162 \\
    9: & ボディ(body): 54 & マスク(mask): 123 & ファン(fan): 58 & 対象(target): 155 \\
    10:& 美女(beautiful girl): 47 & カフェ(cafe): 117 & 声(voice): 56 & 米(US): 148 \\
    11:& 動画(video): 41 & 商品(product): 117 & ラジオ(radio): 52 & 東京都(Tokyo): 143 \\
    12:& 話題(topic): 39 & 春(spring): 108 & ドラマ(drama): 50 & 医療(medical): 139 \\
    13:& 美(beauty): 39 & 味(taste): 107 & 森三中(a comedy group): 49 & 軽症(mild illness): 139 \\
    14:& 美脚(beautiful legs): 39 & 話題(topic): 105 & 妻(wife): 49 & 患者(patient): 134 \\
    15:& 秋元真夏(an idol): 36 & 野菜(vegetables): 105 & 女優(actress): 44 & 職員(staff): 134 \\ 
    16:& コラボ(collaboration): 36 & グッズ(goods): 105 & エール(a TV program): 44 & 病院(hospital): 133 \\
    17:& 水着(swimwear): 35 & 自宅(home): 104 & 日テレ(a TV station): 43 & 企業(company): 114 \\
    18:& オフショット(off-shot): 32 & 絶品(exquisite): 97 & 思い(thought): 43 & 感染拡大(infection spread): 114 \\
    19:& バスト(bust): 32 & 弁当(bento): 93 & 心境(feelings): 43 & 経済(economy): 104 \\
    20:& ビキニ(bikini): 31 & セット(set): 92 & 映画(movie): 42 & 受け入れ(acceptance): 101 \\ \hline
  \end{tabular}
  }
  \label{table:cluster_words_manually}
\end{table*}

\begin{figure}[tb]
    \subfloat[Fitted to user vectors]{\includegraphics[width=0.46\hsize]{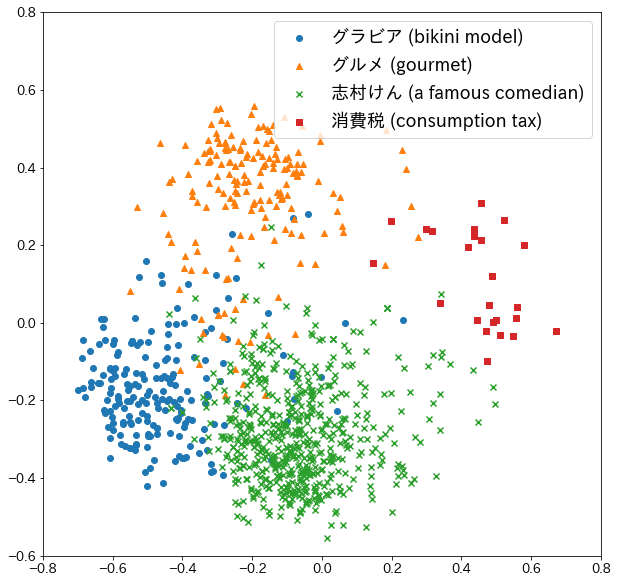}
    \label{fig:pca_user_plot_article_vector_words}}
    \quad
    \subfloat[Fitted to article vector]{\includegraphics[width=0.46\hsize]{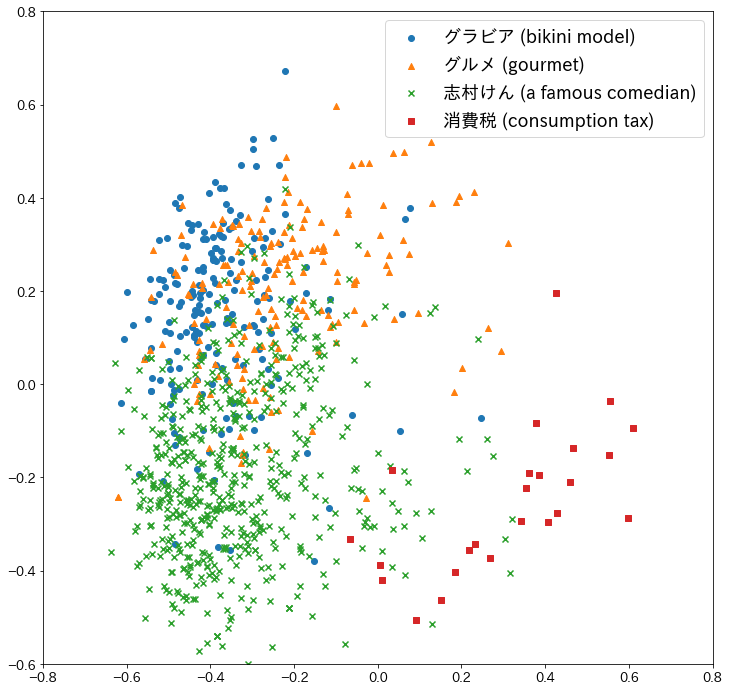}
    \label{fig:pca_article_plot_article_vector_words}}
    \caption{Visualization of the article vectors of specific words. (a) PCA: user, k-means: user, plot: article words. (b) PCA: article, k-means: article, plot: article words.}
\end{figure}

We picked up four words that characterized each region to understand the distribution of the user interest space: ``グラビア(bikini model),'' ``グルメ(gourmet),'' ``志村けん(a famous Japanese comedian who died of COVID-19),'' and ``消費税(consumption tax).''
Fig.~\ref{fig:pca_user_plot_article_vector_words} shows the vector of articles that contains specific words in the user interest space.
All words can be easily identified.
There is an article including ``グラビア(bikini model)'' on the left, one with ``グルメ(gourmet)'' on the top, one with ``志村けん(a famous Japanese comedian)'' on the bottom, and one with ``消費税(consumption tax)'' on the right. 
Interestingly, ``志村けん(a famous Japanese comedian)'' is distributed broadly.
Since he is a comedian, these articles are mainly positioned in $R_C$, but some articles are positioned in $R_D$.
$R_D$ includes ``politics/domestic'' articles, and news about his obituary is strongly correlated to this region.
For comparison with the user interest space, Fig.~\ref{fig:pca_article_plot_article_vector_words} shows the vector of an article that contains a specific word by fitting the PCA to the article vector. 
It can be seen that ``グラビア(bikini model)'' and ``グルメ(gourmet)'' are not distinct.
It was found that features reflecting usage can be obtained from the user interest space.

\subsection{Analysis of Short Dwell Time}\label{sec:analysis}

In this paper, to clarify the significance of short dwell time in relation to user behavior, the following research questions guided the analysis:

\begin{enumerate}
    \item Do users with similar vectors have similar characteristics regarding dwell time?
    \item Is the engagement with content always low with short dwell time?
    \item Is short dwell time meaningful as an element reflecting user interest?
\end{enumerate}

\subsubsection{Do users with similar vectors have similar characteristics regarding dwell time?}

If some users have similar vectors, then they read many similar articles.
We analyzed whether the dwell times of those users had a similar tendency.
If user interest has a similar relation to dwell time, dwell time does not always indicate content quality but it does user interest.

\begin{figure}[tb]
  \begin{center}
  \includegraphics[width=0.85\hsize]{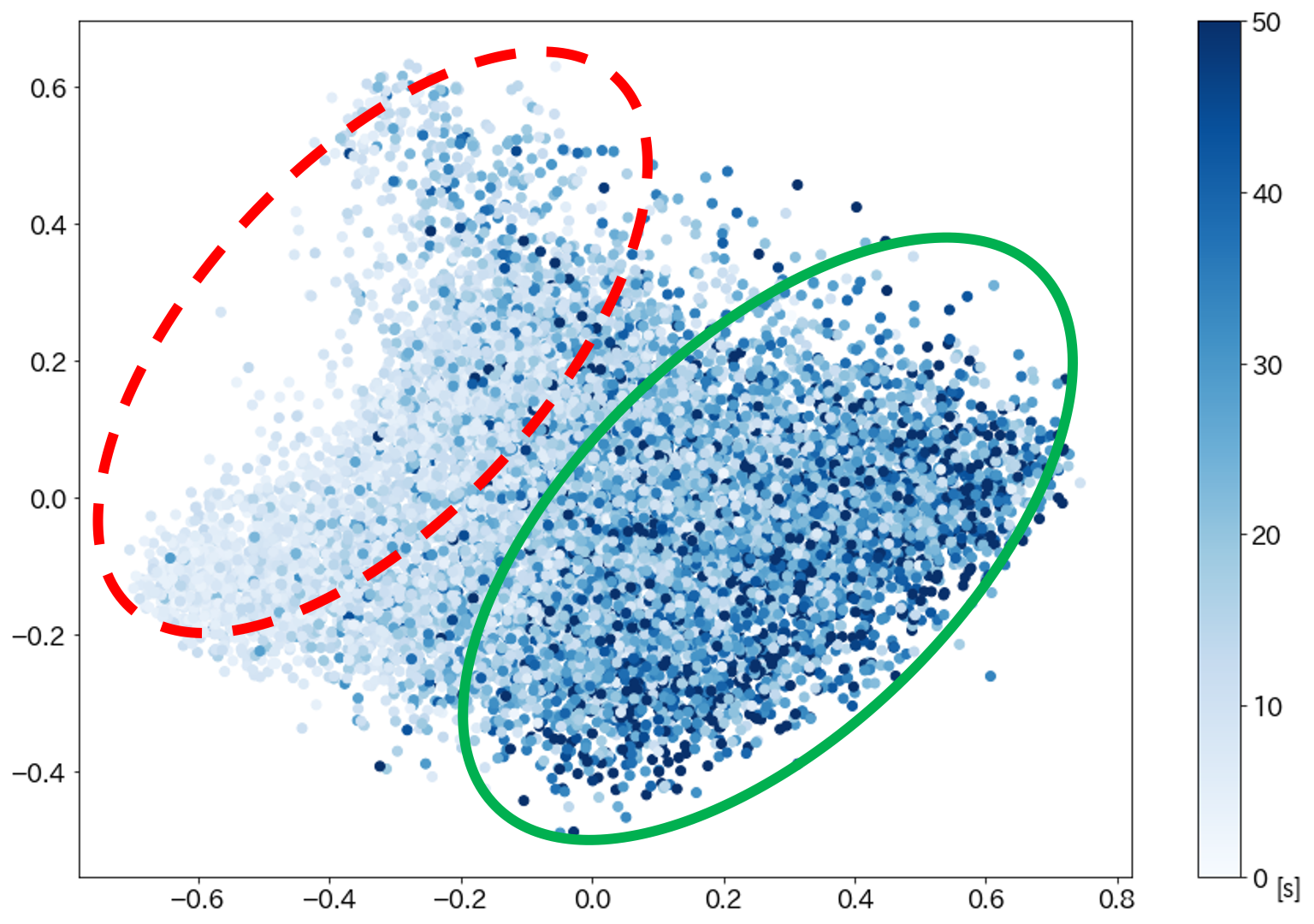}
  \caption{Distribution of the user vector and dwell time. The dwell time tends to be short in the left and top parts (indicated by the red dashed circle) and long in the right and bottom parts (indicated by the green circle).}
  \label{fig:user_vector_dwelltime}
  \end{center}
\end{figure}

Fig.~\ref{fig:user_vector_dwelltime} shows the distribution of the user vector and the dwell time. 
The dwell time is the median value of the user's click history.
The user's dwell time tends to be short in the left and the top parts (indicated by the red dashed circle) of the visualization, and it tends to be long in the right and bottom parts (indicated by the green circle).
The median of users dwell time in each cluster in Fig.~\ref{fig:user_vector} is that the top cluster (blue) is 12.89s, the left cluster (cyan) is 8.11s, the bottom cluster (red) is 22.17s, and the right cluster (green) is 26.66s, and this tendency equals the distribution in Fig.~\ref{fig:user_vector_dwelltime}.
Thus, users with similar vectors have similar characteristics regarding dwell time.

\begin{figure}[tb]
  \begin{center}
  \includegraphics[width=0.9\hsize]{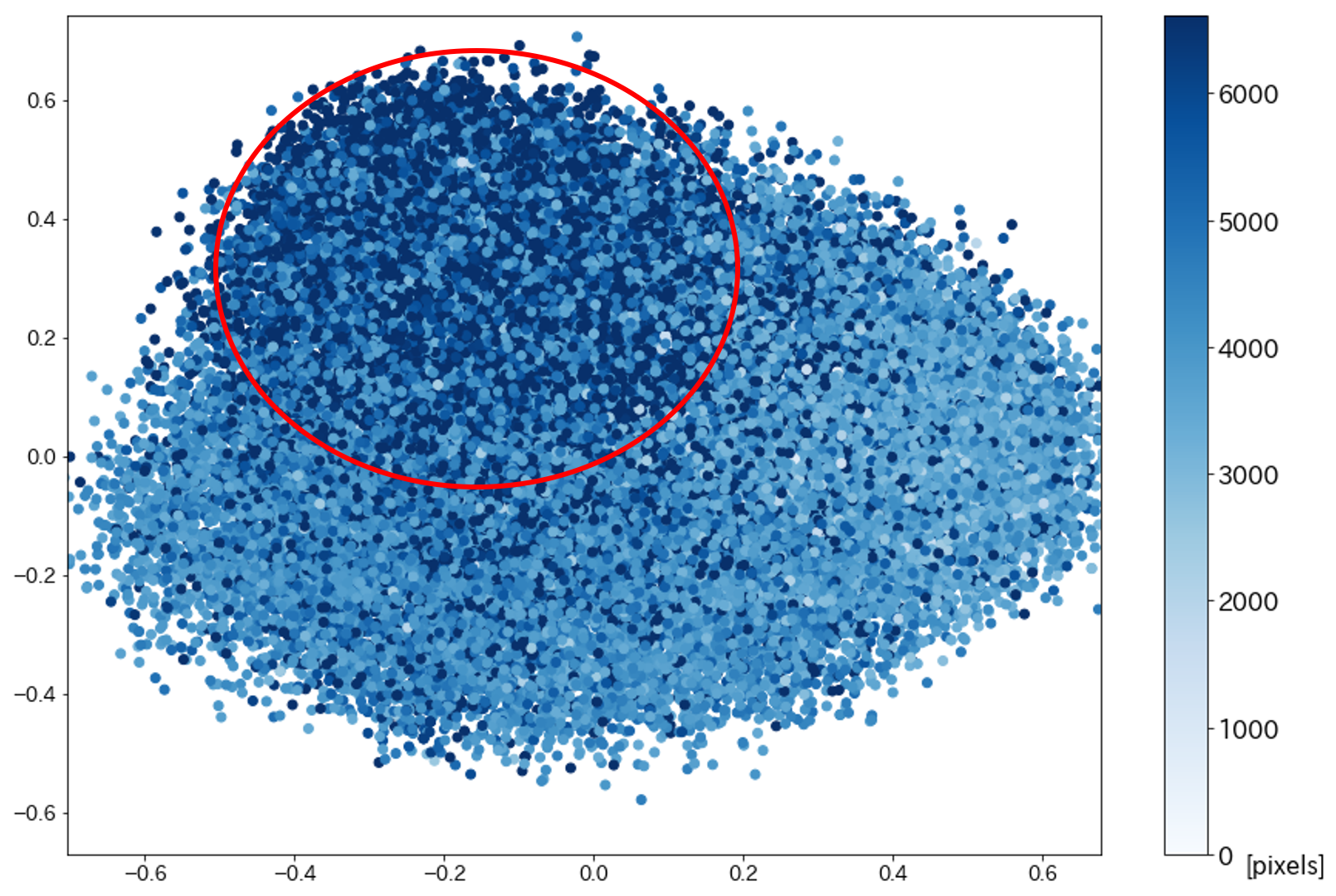}
  \caption{Distribution of the article vector and the article height. The article height tends to be long at the top (indicated by the circle).}
  \label{fig:article_vector_article_length}
  \end{center}
\end{figure}

However, if regions with short dwell time include many articles with a low height, we cannot state that short dwell time is related to the similarity of user characteristics.
Fig.~\ref{fig:article_vector_article_length} shows the distribution of the article vectors and the article height.
As shown in Fig.~\ref{fig:user_vector_dwelltime}, the dwell time tends to be short in the top and left parts, and longer in the right and bottom parts.
Although the article height tends to be long at the top (indicated by the circle), it is at the almost median height in other parts.
Thus, there is no significant correlation between dwell time and article height. 
As shown in Table~\ref{table:cluster_words_manually},  $R_B$ has many words related to ``cooking/lifestyle'' in the title.
Although these articles are long, it can be concluded that the article has not been browsed to the end because the users' dwell time is short.
Thus, regions with short dwell time do not include many articles with low heights, and short dwell time is related to the similarity of user characteristics.

\subsubsection{Is the engagement with content always low with short dwell time?}

To discuss engagement, we revisit in Fig.~\ref{fig:user_vector_dwelltime}.
Many users in the left part ($R_A$) and the top part ($R_B$) of the figure had a short dwell time.
If users with short dwell time complained of an application, they withdrew the application immediately.
However, the users included in this dataset were continuously using this service, so their engagement was not low despite a short dwell time.
We consider that $R_A$ users, who frequently read ``female idol group/bikini model'' articles, were satisfied by seeing only the images without reading the text component of the content; thus, their dwell time was comparatively short.
Overall, we conclude that these users with short dwell time were satisfied with the application.

\subsubsection{Is short dwell time meaningful as an element reflecting user interest?}

To answer this research question, we removed short dwell time articles from user click history.
If clicked articles with short dwell time were not especially significant for user interest, the user vector was not significantly changed by excluding these articles from user click history.
However, if the user vector excluding these articles significantly differed from the original user vector, short dwell time could be considered meaningful as an element expressing user interest.

To analyze the characteristics of short dwell time browsing, we define two types of user vectors.
A user vector created by excluding clicks with short dwell time is defined as $u_{long}$.
We define $u_{short}$ as a user vector created with short dwell times only.
$A_{u,short}$ describes a set of news articles that are clicked on by $u$ with a dwell time of 5s or less, and $A_{u,long}$ is a news article clicked on by $u$ and viewed for more than 5s. 
Each term is defined as follows:

\begin{equation}
A_u = A_{u, short} \cup A_{u, long},
\end{equation}

\begin{equation}
u_{long}:=\frac{\sum_{a \in A_{u,long}} a}{\left\|\sum_{a \in A_{u,long}} a\right\|} \in \mathbb{R}^{d},
\end{equation}

\begin{equation}
u_{short}:=\frac{\sum_{a \in A_{u,short}} a}{\left\|\sum_{a \in A_{u,short}} a\right\|} \in \mathbb{R}^{d}.
\end{equation}

\begin{figure}[tb]
  \centering
  \includegraphics[width=0.9\hsize]{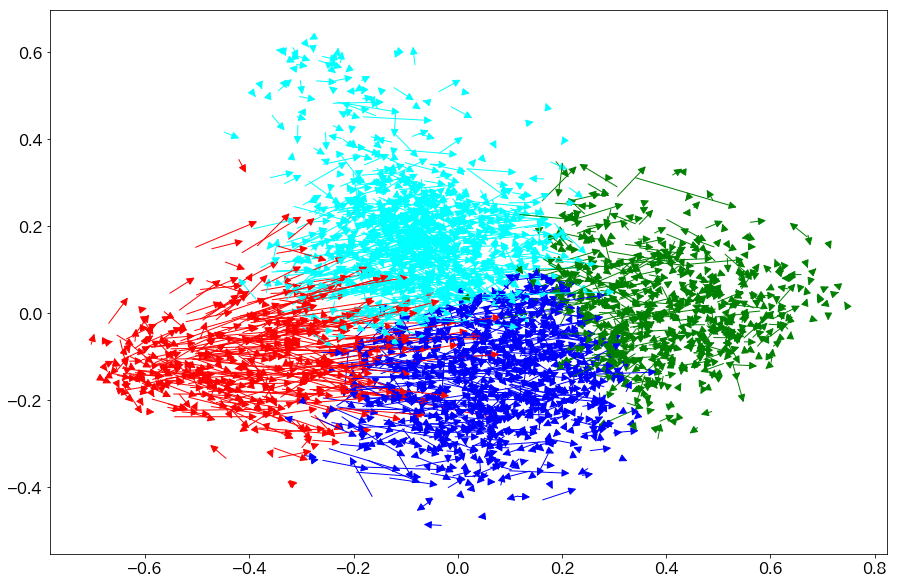}
  \caption{Changes in the user vector when browsing sessions with a dwell time of 5s or less are excluded. Arrows show user vectors after excluding clicks with short dwell time. 77.85\% of user vectors moved to the right.}
  \label{fig:user_vector_long}
\end{figure}

Fig.~\ref{fig:user_vector_long} shows the change after excluding clicks with short dwell time.
For ease of viewing, we randomly sampled 3,000 users (a quarter of the total users).
Arrows show user vectors after excluding clicks with short dwell time and 77.85\% of user vectors moved to the right.
It can be seen that there are many articles with short dwell time on the left side of the user interest space.
This change between user vectors means that the user browsed articles with different preferences depending on whether the dwell time was short.

To understand this change, we analyze these two user vectors using clustering.
Note that long and short fit into different k-means models.
Each long cluster is indicated by $A$, $B$, $C$, and $D$, and each short cluster is indicated by $A'$, $B'$, $C'$, and $D'$. 
The corresponding clusters of symbols are the closest to each other (e.g., $A$ and $A'$).
To identify the closest cluster, we define the distance between cluster $i$ and cluster $j$ as follows:

\begin{equation}
|| {\rm centorid}_{i} - {\rm centroid}_{j} ||.
\end{equation}

\begin{figure}[tb]
  \centering
  \subfloat[Long]{\includegraphics[width=0.8\hsize]{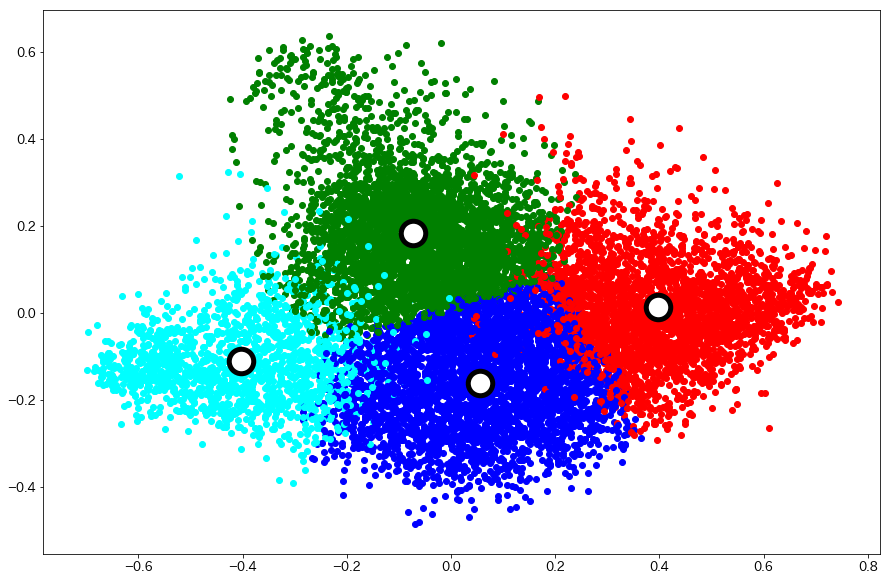}\label{fig:user_vector_new_centroid_long}}\quad
  \subfloat[Short]{\includegraphics[width=0.8\hsize]{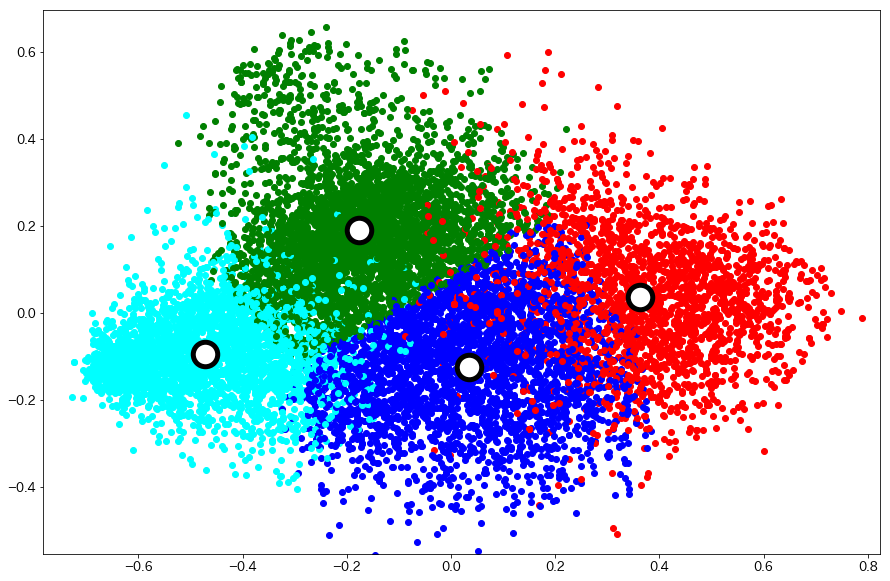}\label{fig:user_vector_new_centroid_short}}
  \caption{User clusters and centroids. At short, where the dwell time is short, the user vector moves to the left.}
  \label{fig:user_vector_new_centroid}
\end{figure}

Fig.~\ref{fig:user_vector_new_centroid} shows the user cluster and centroid of each user cluster. 
The tendency of clustering is not significantly changed, but in comparing the centroids of long and short, these centroids are comparatively located to the left in short.
This move is similar to that shown in Fig.~\ref{fig:user_vector_long}.

If there are some users whose cluster change between long and short, their short dwell time would have a strong tendency to express user interest.
Table~\ref{table:user_vector_new_centroid} shows the number of users moving from long to short and the distance in each cluster.
The distance from cluster $D$ to $A'$ is 0.43 and the number of users moving is 26, whereas the distance from clusters $A$ to $D'$ is 0.54 and the number of users moving is 851.
From this, it can be concluded that the distance and the number of moving users are not proportional.
Surprisingly, excluding short dwell time click history, 30.87\% of user clusters changed significantly between the two vectors.
Thus, by excluding short dwell time click history, the user vector significantly changes in the user interest space.

\begin{table}[tb]
  \caption{The number of users moving from long to short and the distance of centroids in each cluster. A: Blue, B: Green, C: Red, D: Cyan. Bold represents the clusters with the closest distance of the centroid.}
  \centering
  \scalebox{0.75}[0.9]{
      \begin{tabular}{|c|c|c|c|c|c|c|c|c|} \hline
        & \multicolumn{4}{c|}{\# of moves} & \multicolumn{4}{c|}{distance} \\ \hline
        \backslashbox{long}{short} & $A'$ & $B'$ & $C'$ & $D'$ & $A'$ & $B'$ & $C'$ & $D'$ \\ \hline
        $A$ & \textbf{2065} & 431 & 348 & 851 & \textbf{0.041} & 0.41 & 0.36 & 0.53  \\
        $B$ & 258 & \textbf{3064} & 113 & 466 & 0.32 & \textbf{0.10} & 0.46 & 0.48 \\
        $C$ & 793 & 214 & \textbf{1659} & 129 & 0.38 & 0.60 & \textbf{0.04} & 0.87 \\
        $D$ & 26  & 35 & 10 & \textbf{1437} & 0.43 & 0.37 & 0.77 & \textbf{0.07} \\ \hline
      \end{tabular}
  }
  \label{table:user_vector_new_centroid}
  \vspace{-4mm}
\end{table}

\begin{figure}[tb]
  \begin{center}
  \includegraphics[width=0.75\hsize]{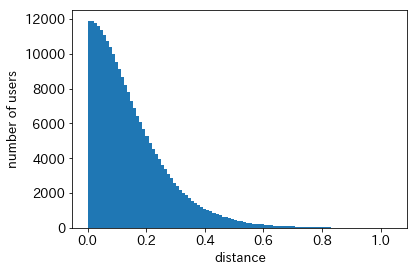}
  \caption{Long to short distance cumulative histogram. The x-axis indicates the distance. The average is 0.2 and the median is 0.18.}
  \label{fig:user_vector_long_to_short_distance_distribution_cumulative}
  \end{center}
\end{figure}

From the above results, the difference of user interest between long and short has been shown, and now we attempt to quantitatively determine how different they are.
Fig.~\ref{fig:user_vector_long_to_short_distance_distribution_cumulative} shows the cumulative histogram of the distance between long and short.
This distance for a user is:
\begin{equation}
|| u_{long} - u_{short} ||.
\end{equation}
The average value is 0.2 and the median is 0.18.
This result shows that some users have different interests between long and short vectors.
In addition, several users have a large distance between long and short and they have very different interests.
Overall, long and short show different characteristics, and there is a relationship between user interest and dwell time.

\subsection{Results Summary}

The following results were obtained for the research questions presented in this section.
We found these results to visualize articles and users on our user interest space.

\begin{enumerate}
    \item Users with similar vectors showed that they also had similar characteristics regarding dwell time.
    \item Engagement with content is not always low when dwell time is short.
    \item It was shown that short dwell time is meaningful as an element expressing user interest.
\end{enumerate}

From the above results, we conclude that short dwell time does not always indicate low user engagement; it can also express user interest.

\section{Another Period Analysis}
In this section, we confirm that these results are valid for different time periods.
Our results are based on the visualization of user interest space, so we verified whether this space could achieve interpretable visualization even with data from different periods.
We also confirmed that the results of the dwell time analysis show a similar tendency, but due to limited space, we cannot describe this in detail.

For this study, we primarily collected two weeks of data from April 2 to April 15, 2020.
We performed a similar analysis of one month of data for November 2019 to ensure the validity of the analysis.
The results of the second analysis are similar to the results we have shown thus far.
We call this dataset Nov-2019 and the previous dataset Apr-2020.
However, there were also several differences. 
Users in Nov-2019 was sampled in the same way as users in Apr-2020 (found in Section \ref{sec:data}).

\begin{figure}[tb]
  \begin{center}
  \includegraphics[width=0.7\hsize]{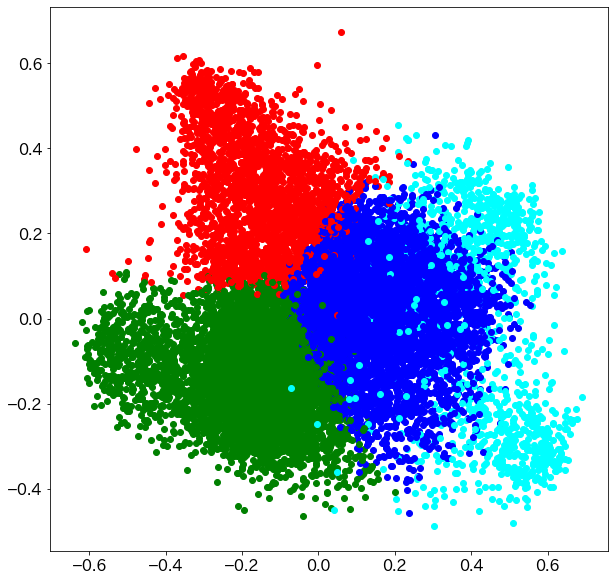}
  \caption{The user vector for the Nov-19 dataset (k-means, k=4). Unlike in Fig.~\ref{fig:user_vector}, it is divided into only three directions.}
  \label{fig:user_vector_other_period}
  \end{center}
\end{figure}

Fig.~\ref{fig:user_vector_other_period} shows the user vector in Nov-2019.
Note that the k-means and PCA space in Nov-2019 and Apr-2020 do not correspond; that is, the colors do not correspond to each other because k-means assigned each cluster a color randomly.
The user vector was divided into four types of interest in Apr-2020, but only three types of interest in Nov-2019.

\begin{figure}[tb]
  \begin{center}
    \includegraphics[width=0.9\hsize]{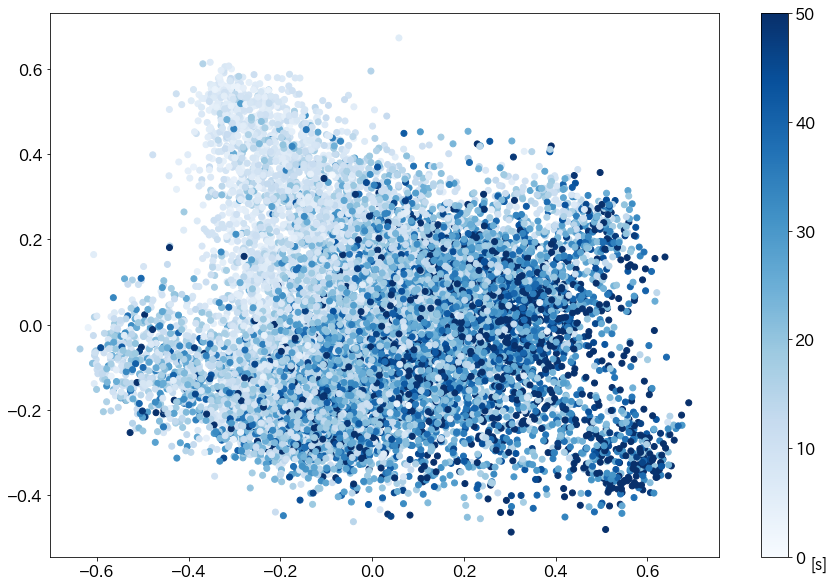}
    \caption{Distribution of the user vector and dwell time in Nov-2020.}
    \label{fig:user_vector_dwelltime_another_period}
    \end{center}
\end{figure}

Fig.~\ref{fig:user_vector_dwelltime_another_period} shows the distribution of user dwell time in user interest space in Nov-2019.
As in Apr-2020, users with short dwell times were crowded in a specific position in this space.
This result shows that short dwell time is related to user interest across different periods of time.\footnote{
Of course, by the difference of periods, the meaning of vector spaces and clusters were changed. However we consider that this change was not affected for our contribution and answers of our research questions.
}

\label{sec:another_period}

\section{Conclusion}
In this study, we conducted various analyses to clarify what short dwell time means for user interest.
Many previous studies presuppose that content with short dwell time indicates low quality or low user engagement.
However, we hypothesized that the short dwell time does not always indicate low content quality or low user engagement, but is instead related to user interest.
Therefore, we analyzed short dwell time browsing in detail using user behavior logs from a large-scale mobile news application.

In future work, we plan to investigate how the recommendation list changes when browsing with a short dwell time is excluded in an actual recommender system.
Although this study adopts conventional methods, we try to use some novel methods, such as BERT and t-SNE. 
Furthermore, we leave new methods and/or parameters can improve user satisfaction by taking both short and long dwell times into account in recommender systems on the basis of this future work.
\label{sec:conclusion}

\bibliographystyle{IEEEtran}
\bibliography{references.bib}

\end{document}